\documentclass[11pt]{article}
\usepackage{format}
\usepackage[margin=1in]{geometry}

\def\T{{ \mathrm{\scriptscriptstyle T} }}
\begin{document}
\setlength{\abovedisplayskip}{5pt}
\setlength{\belowdisplayskip}{5pt}
\setlength{\abovedisplayshortskip}{5pt}
\setlength{\belowdisplayshortskip}{5pt}
\title{Multiplicative Effect Modeling: The General Case}
\author{Jiaqi Yin\thanks {Department of Biostatistics, University of Washington, Seattle, WA 98195, USA. Mail to: jiaqiyin@yahoo.com}
\and Sonia Markes \thanks{Department of Statistical Sciences, University of Toronto, Toronto, Ontario M5S 3G3, Canada. Mail to: so-
nia.markes@mail.utoronto.ca}
\and Thomas S. Richardson \thanks{Department of Statistics, University of Washington, Box 354322, Washington 98195, U.S.A.  Mail to: thomasr@u.washington.edu}
\and Linbo Wang \thanks{Department of Statistical Sciences, University of Toronto, Toronto, Ontario M5S 3G3, Canada. Mail to: linbo.wang@utoronto.ca}}

\date{}
\maketitle

\begin{abstract}
   Generalized linear models, such as logistic regression, are widely used to model the association between a treatment and a binary outcome as a function of baseline covariates. However, the coefficients of a logistic regression model correspond to log odds ratios, while  subject-matter scientists are often interested in relative risks. 
   Although odds ratios are sometimes used to approximate relative risks, 
   this approximation is appropriate only when the outcome of interest is rare for all levels of the covariates.   Poisson regressions do measure multiplicative treatment effects including relative risks, but with a binary outcome not all combinations of parameters lead to fitted means that are between zero and one. Enforcing this constraint makes the parameters variation dependent, which is undesirable for modeling, estimation and computation. Focusing on the special case where the treatment is also binary, \cite{Richardson2017} propose a novel binomial regression model, that allows direct modeling of the relative risk. The model uses a log odds-product nuisance model leading to variation independent parameter spaces. Building on this we present general approaches to modeling the multiplicative effect of a continuous or  categorical treatment on a binary outcome.  
   Monte Carlo simulations demonstrate the desirable performance of our proposed methods. A data analysis further exemplifies our methods.
\end{abstract}
\textbf{Keywords:}
Binomial Regression Model; Generalized Odds Product; Multiplicative Treatment Effect.

	\section{Introduction}

The logistic model is widely used to model binary outcomes, such as alive versus dead, yes
versus no, success versus failure, and so on. In a logistic model,  the probability of the outcome
$Y$ is modeled as a function of covariates using a logit function. The coefficient associated with 
a particular binary covariate, which we will refer to as treatment, is a log-odds ratio.
Here the odds is simply the ratio between the probability of  $Y=1$ and the probability of $Y=0$, and an odds ratio is the ratio between the odds for two different levels of treatment.
Since the resulting likelihood is concave, it is  feasible to compute maximum likelihood estimates for large data sets.

However, in many epidemiological and other studies,  researchers are primarily
interested in estimating the effect of a treatment $Z$ on a binary outcome $Y$ on the multiplicative scale  \citep{Lumley2006}. 
Specifically, this can be formulated as a series of relative risks, which are ratios contrasting the probability of $Y=1$ in treatment group $Z = z$ versus the probability of $Y=1$ in a baseline group $Z = z_0$:
$
\rr(z_0, z) = {\text{pr}(Y=1\mid Z=z)}/{\text{pr}(Y=1 \mid Z = z_0)}.
$
In this paper, we consider a continuous or categorical treatment $Z$. 

In practice odds ratios are sometimes used to approximate relative risks. However, when the outcome is prevalent, odds ratios and relative risks may be very different. 
Consequently, it is not usually appropriate to use odds ratios as an approximation for relative risks when the latter is of primary interest. There are also qualitative differences between these measures: whereas relative risks are collapsible,
so that the marginal relative risk will
lie in the convex hull of stratum-specific relative risks  \citep{greenland1999}, the same does not hold for odds ratios.





Within the generalized linear model framework, multiplicative treatment effects are typically modeled via a Poisson regression which imposes a linear association between the log of the probability of $Y=1$ and covariates. However, Poisson regression with a binary outcome has drawbacks in terms of modeling, prediction and computation. This is because $\rr(z_0,z)$ is variation dependent on the baseline probability $\pr(Y=1\mid Z=z_0)$. For example, if $\rr(z_0,z)=2$, then $\pr(Y=1\mid Z=z)=2\times \pr(Y=1\mid Z=z_0)$, so that $\pr(Y=1\mid Z=z_0)\leq 0.5$. 
Therefore there is a restricted domain over which the quantities $\left[\{\rr(z_0,z);z\},\pr(Y=1\mid Z=z_0)\right]$ are compatible with a valid probability distribution.
This may lead to misspecification when modeling. Also the fitted probability for any treatment given covariates can go outside of the range $[0,1]$.

\cite{Richardson2017} provide a simple alternative approach to modeling the relative risk. Specifically, in addition to specifying a model for $\log (\rr)$, they propose a  nuisance model for the log of the odds product (\textsc{op}):
$$
\op (z_0,z)= \frac{\pr(Y=1\mid Z=z)\pr(Y=1\mid Z=z_0)}{\{1 - \pr(Y=1\mid Z=z)\}\{1 - \pr(Y=1\mid Z=z_0)\}}.
$$
This leads to an unrestricted domain for which the quantities $\left[\{\rr(z_0,z);z\},\op(z_0,z)\right]$ are compatible with a valid probability distribution.
Consequently, with the odds product model, the relative risk can be
estimated via either unconstrained maximum likelihood estimation or semi-parametric g-estimation methods.
However, their method is restricted to binary treatments.

\cite{Tchetgen2013} and \cite{Dukes2018} approach the problem of estimating relative risks by providing alternative nuisance models. In contrast
to the choice of \cite{Richardson2017}, their choices of nuisance models apply to both binary and continuous treatments, but they do not lead to the full likelihood. Consequently, in their proposals, the
relative risk is estimated with g-estimation methods, but not maximum likelihood estimation.

Building on \cite{Richardson2017}, we present two new approaches that model multiplicative effects with continuous or categorical treatments.  
The first imposes an assumption that the relative risk is a monotone function of an ordinal treatment. The second introduces a 
new nuisance model, using a so-called generalized odds product \citep{wang2017congenial}, that is variation independent of the relative risks.


\section{Methodology}
\label{sec:method}
\subsection{Parameterization with Monotonic Treatment Effects}\label{sec:mono}



Denote the relative risk between a treatment $z$ and the baseline treatment  $z_0$ given a random vector of covariates $v$ as
$\rr(z_0, z; v) = {\pr(Y=1\mid V=v, Z = z)}/{\pr(Y=1\mid V=v, Z = z_0)},$
where $Z$ can be  continuous or categorical.
For notational simplicity, we denote  $\pr(Y=1\mid Z=z, V=v)$ as $p_z(v)$,
the odds product of treatment $z$ and baseline treatment $z_0$ as
$\op(z_0,z; v) = {p_0(v)p_z(v)}/[{\{1-p_0(v)\}\{1-p_z(v)\}}].$


To fix ideas, first consider the special case where $Z$ is a continuous treatment taking values in a bounded interval, say $[0,1].$ Our goal is to find $\phi(v)$ so that  for any $v$,
the mapping given by 
$$
\left[\log \{\rr(0,z;v)\}, z\in [0,1]; \phi(v)\right] \rightarrow \left\{p_z(v), z\in [0,1]\right\}
$$
is a diffeomorphism between the interior of their domains.
\cite{Richardson2017} show that if we let $\phi(v) = \log \{\op(0,1;v)\}$, then any possible value of $[\log\{\rr(0,1;v)\}, \phi(v)]$ implies that $\{p_0(v), p_1(v)\}\in (0,1)^2.$ The key insight for our development is that if the log relative risk $\log\{\rr(0,z;v)\}$ is monotonic in $z$ for all covariate values $v$, then $p_z(v)$ is also monotonic in $z$.
Consequently, 
$$0 < \min\{p_0(v),p_1(v)\} \leq p_z(v) \leq \max\{p_0(v),p_1(v)\}< 1 \quad (z\in [0,1]).$$
Therefore, any possible value of $[\log \{\rr(0,z;v)\},\phi(v)]$ such that $\log\{\rr(0,z;v)\}$ is monotone in $z$ implies that $p_z(v) \in (0,1)$ for all $z\in[0,1]$.


The monotonic treatment effect assumption we have invoked may be considered reasonable in many real-life situations. For example, the recovery probability in the arm receiving full-dosage is usually at least as high as in the small-dosage arm \citep{mamgani2008}, and greater income may be associated with a higher probability of satisfaction  \citep{richard2001}.


This idea above can be generalized to accommodate more types of variables for the treatment $Z$, including ordinal and unbounded continuous variables. 




\begin{theorem}[Variation independence with monotonic treatment effects]
	\label{thm:main}
	Let $\mathcal{Z}\subseteq \mathbb{R}$ and $\mathcal{V}$ be the support of $Z$ and $V$, respectively. Let $h(z,v)$ and $g(v)$ be real-valued functions with support $\mathcal{Z} \times  \mathcal{V} $ and $\mathcal{V}$, respectively. 
	If $h(z,v)$ is bounded in $z$ and for each $v$, monotonic in $z$, then
	there exists a unique set of proper probability distributions $\{p_z(v); z\in \mathcal{Z}, v\in \mathcal{V}\}$ 
	such that $\log\{\rr(z_0,z;v)\}=h(z,v)$  and $\log\{\op(z_
	{\inf},z_{\sup};v)\}=g(v)$, where $z_{\inf} = \inf \{z:z\in \mathcal{Z}\}$, $z_{\sup} = \sup\{z:z\in \mathcal{Z}\}$ and 
	$$
	\op(z_
	{\inf},z_{\sup};v) = \mathop{\lim}\limits_{z_1\rightarrow z_{\inf}}\lim\limits_{z_2\rightarrow z_{\sup}} \dfrac{p_{z_1}(v)p_{z_2}(v)}{\{1-p_{z_1}(v)\}\{1-p_{z_2}(v)\}}.
	$$
\end{theorem}

\begin{remark}
	The boundedness condition on $h(z,v)$ guarantees that the implied probabilities $p_z(v)$ are bounded away from 0. 
\end{remark}





In our simulations and data analysis, we consider a bounded treatment $Z$ and the following models for
$\log\{\rr(z_0,z;v)\}$  
and $\log\{\op(z_{\min},z_{\max};v)\}$:
\begin{align}
	\log\{\rr(z_0,z; V, \gamma)\} &= \gamma^\T  V (z-z_0) \quad z\in \mathcal{Z}, \label{eq:model of rr in mono}\\
	\log\{\op(z_{\min},z_{\max}; V, \beta)\} &= \beta^\T V  \label{eq:model of op in mono},
\end{align}
where $z_{\min} = \min \{z:z\in \mathcal{Z}\}$, $z_{\max} = \max\{z:z\in \mathcal{Z}\}$.
In light of the boundedness condition on $h(z,v)$, when the treatment is unbounded, the specification 
\eqref{eq:model of rr in mono} may be modified to:
\begin{align}
	\log\{\rr(z_0,z; V, \gamma)\} &= \gamma^\T  V f(z) \quad z\in \mathcal{Z}, \label{eq:model of rr in mono unbounded z}
\end{align}
where $f(\cdot)$ is a  bounded function.

The log-likelihood for a unit $i$  can be written as
\begin{equation*}\label{eq:log-likehood in mono}
	l(\gamma,\beta|z_i,v_i,y_i) =  y_i\log\{p_{z_i}(v_i;\gamma, \beta)\} +  (1-y_i)\log\{1-p_{z_i}(v_i;\gamma, \beta)\}.
\end{equation*}
Inference for  $\gamma$ and $\beta$ can be performed in the standard way. We provide explicit formula for Wald-type confidence intervals in the Supplementary Material. 


We finish this part with a remark that the proposed assumption of monotonic treatment effects may be falsified from the observed data. In practice, analysts may use descriptive plots to examine the relationship between the treatment and outcome, and use them to assess the plausibility of the monotonic treatment effect assumption. See \S\ref{sec:application} for an illustration and the Supplementary Material for simulations of violations of the monotonicity condition.

\subsection{Parameterization with a categorical treatment}\label{sec:gop}

The approach introduced in \S\ref{sec:mono} is not directly applicable if the relative risk is not monotonic in $z$.
We now consider a categorical treatment whose effect on the outcome is not necessarily monotonic. Suppose that the treatment $Z$ takes values in  $\{z_0,\ldots, z_K\}$, where $z_0$ is chosen as the baseline treatment. The quantities of interest are relative risks  $\rr(z_0,z_k;v)\ (k=1,\ldots,K)$. 
For notational simplicity, we denote  $\pr(Y=1\mid Z=z_k, V=v)$ as $p_k(v)$, and $\rr(z_0,z_k;v)$ as $\rr(0,k;v)$. Following \cite{wang2017congenial}, we introduce a nuisance model on  the generalized odds product 
\begin{equation*}
	\gop(v) = \prod_{k=0}^K \frac{p_k(v)}{1-p_k(v)}.
\end{equation*}
The following theorem states that the generalized odds product is variation independent of the set of relative risks. 

\begin{theorem}[Variation independence with a categorical treatment]\label{thm:gop-variation-indep}
	Let $\mathcal{M}$ denote a $(K+1)$-dimensional model on 
	\begin{align*}
		\rr(0,k;v) &= \frac{p_k(v)}{p_0(v)}\quad (k=1,\ldots,K),\\
		\gop(v) &= \prod_{k=0}^K \frac{p_k(v)}{1-p_k(v)}.
	\end{align*}
	For any $v$, the mapping given by
	\begin{equation*}
		\left\{p_0(v),\ldots,p_K(v)\right\} \rightarrow \left[\log \{\rr(0,1;v)\}, \ldots, \log \{\rr(0,K;v)\},\log \{\gop(v)\}\right] \label{map:p-to-model}
	\end{equation*}
	is a diffeomorphism from $(0,1)^{K+1}$ to $(\R)^{K+1}$. Furthermore, the models in $\mathcal M$ are variation independent of each other.
\end{theorem}
The proof of Theorem \ref{thm:gop-variation-indep} is deferred to the Supplementary Material. In our simulations and data analysis, we consider the following specifications of $\mathcal{M}$:
\begin{align}
	\log \{\rr(0,k;v)\} &= \alpha_k^\T X \quad (k=1,\ldots,K)\label{eq:gop-rr},\\
	\log \{\gop(v)\} &= \beta^\T W,\label{eq:gop-gop}
\end{align}
where $X=X(v)$, $W=W(v)$. Theorem \ref{thm:gop-variation-indep} shows that the parameters $\alpha_1,\ldots, \alpha_K$, and $\beta$ are variation independent so that their domains are unconstrained. Maximum likelihood estimates and associated inference for
parameters $\alpha_1,\ldots, \alpha_K$, and $\beta$ can then be obtained in standard fashion. 
The relative risk model  in this approach is more flexible than the corresponding model \eqref{eq:model of rr in mono} in \S\ref{sec:mono}, which assumes monotonicity, thus \eqref{eq:gop-rr} has $K$-times as many  parameters.

\begin{remark}
	In general the log-likelihoods given by \eqref{eq:model of op in mono} and \eqref{eq:model of rr in mono unbounded z}  or by 
	\eqref{eq:gop-rr} and \eqref{eq:gop-gop} may not be concave. In the Supplementary Material we describe
	a simple iterative procedure for finding a solution to the score equations. Specifically, for the method in \S \ref{sec:mono} we iterate between $\beta$ and $\gamma$ to maximize the likelihood, while for the method of \S \ref{sec:gop}  we iterate sequentially among $\alpha_1, \ldots, \alpha_K, \beta$.  {It has been shown that this iterative partial maximization algorithm increases the likelihood at each step and hence will converge to a stationary point \citep[][Appendix, Proposition 1]{drton:eichler:2006}.}
\end{remark}


\section{Simulation}\label{sec:simulation}

\begin{table}[! htbp]
	\centering
	\begin{threeparttable}
		\def~{\hphantom{$-$}}
		\caption{Simulation results for three different methods based on 500 samples and 1000 Monte Carlo runs. The true values for $\gamma$, $\alpha_1$ and $\alpha_2$ are $(0,1)^\T$, $(0,1)^\T$ and $(0,2)^\T$ respectively.}
		{%
			\begin{tabular}{lcccccccccc}
			\hline
		& \multicolumn{2}{c}{Bias${\scriptscriptstyle \times 10^2}$(Standard Error${\scriptscriptstyle \times 10^2}$)} && \multicolumn{2}{c}{SD Accuracy} && \multicolumn{4}{c}{Coverage  (Nominal = 95\%)}
		\\[2pt] 
		\midrule
		Setting I & \multicolumn{2}{c}{$\gamma$}   & & \multicolumn{2}{c}{$\gamma$}  & &&\multicolumn{2}{c}{$\gamma$}& \\[3pt]
		Monotone & \multicolumn{2}{c}{~0.232(0.214)} & & \multicolumn{2}{c}{0.996}   & &&\multicolumn{2}{c}{0.950} & \\ 
		& \multicolumn{2}{c}{~0.442(0.256)} & &
		\multicolumn{2}{c}{1.011}  & 
		&&\multicolumn{2}{c}{0.958} & \\[5pt] 
		DR-G & \multicolumn{2}{c}{$-$0.267(0.763)} & & \multicolumn{2}{c}{0.659}  & 
		&&\multicolumn{2}{c}{0.890}& \\ 
		&\multicolumn{2}{c}{~20.93(1.250)}  &&
		\multicolumn{2}{c}{0.558}  & 
		&&\multicolumn{2}{c}{0.816}& \\[6pt]
		\midrule
		Setting II &  $\alpha_1$ & $\alpha_2$ && $\alpha_1$& $\alpha_2$ &&& $\alpha_1$ & $\alpha_2$ &\\[3pt]
		GOP &  $-$0.327(0.495) & 0.788(0.489) && 1.008 & 1.004  &&& 0.958 & 0.955 &\\ 
		&  ~0.462(0.596) & 0.178(0.565) && 1.005 & 1.004 &&& 0.956 & 0.952 & \\[3pt]
			\hline
		\end{tabular}}
		\begin{tablenotes}\footnotesize
		\item Monotone, using models \eqref{eq:model of rr in mono} and \eqref{eq:model of op in mono}; DR-G, doubly robust estimator by \cite{Dukes2018}; GOP, using models  \eqref{eq:gop-rr} and \eqref{eq:gop-gop}.
		\item SD Accuracy = estimated standard deviation / Monte Carlo standard deviation.
		\end{tablenotes}
		\label{tab:sim-all}
	\end{threeparttable}
\end{table}

	We use the following generating model:  Treatment $Z$ is assigned according to a multinomial logistic regression model such that 
\begin{align}
	\label{eqn:ps}
	\log \left\{\frac{\pr(Z=1\mid V)}{\pr(Z=0\mid V)} \right\}&= \eta_1^\T V,\quad\hbox{ and }\quad
	\log \left\{\frac{\pr(Z=2\mid V)}{\pr(Z=0\mid V)} \right\}\;=\; \eta_2^\T V,
\end{align}
where $\eta_1  = (1, -1)^\T$, $\eta_2 = (1, -2)^\T$. 
The covariate vector $V$ includes an intercept and a draw from a uniform distribution on $[-2,2]$.

We consider two different settings for the outcome $Y$:
In Setting I, we treat $Z$ as continuous and $Y$ is generated according to models \eqref{eq:model of rr in mono} and \eqref{eq:model of op in mono}, where $\gamma = (0,1)^\T$, $\beta=(1, -0.5)^\T$, so that the log relative risk is linear in $z$. We apply the method of \S \ref{sec:mono} to estimate the relative risk in this setting, and 
compare it to the doubly robust g-estimator of \cite{Dukes2018}.
In Setting II, $Z$ is viewed as unordered categorical 
and the outcome $Y$ is generated from models \eqref{eq:gop-rr} and \eqref{eq:gop-gop}, where $\alpha_1=\gamma=(0,1)^\T$, $\alpha_2=(0, 2)^\T$, $\beta=(1, -0.5)^\T$. Here, we  apply the method of \S \ref{sec:gop} only.


Table \ref{tab:sim-all} summarizes the simulation results for sample size $500$. 
The bias of our proposed estimators is small when the sample size is 500, and further decreases as the sample size increases; see Table \ref{tab:sim-mono}  in the Supplementary Material. The standard deviation accuracy, defined as the ratio of estimated standard deviation and  Monte Carlo standard deviation, is close to $1$ for our proposed estimators. 
The coverage probability of the proposed Wald-type confidence intervals also achieve the nominal 95\% coverage-rate.  Even though, in theory, the doubly robust g-estimator is consistent in this setting as the propensity score model is correctly specified, as shown in Table \ref{tab:sim-all}, with a small sample of 500 the bias can be very large relative to the standard error. In this case, the model-based standard deviation estimate is also much smaller than the Monte Carlo standard deviation. 
One can also see that in this simulation, the proposed estimator is much more efficient than the g-estimator. {In \S\ref{supple:simu-mis-op} of the Supplementary Material, we consider an alternative setting where the odds product model \eqref{eq:model of op in mono} is mis-specified. In this case, when the sample size is large enough, the doubly robust g-estimator of \cite{Dukes2018} has small bias and nominal coverage while the proposed estimator does not.
	However, in the same setting, when the sample size is small, the doubly robust g-estimator has large bias, relative to the  method proposed in \S\ref{sec:mono}}.

{We further note that our method in \S\ref{sec:mono} assumes that treatment effects are monotonic in $z$ for all levels of $v$. In \S\ref{supple:violation-monotonicity} of the Supplementary Material, we conduct additional simulations to evaluate the performance of the method of \S \ref{sec:mono} when this assumption is violated. }

With a categorical treatment taking $K$ + 1 levels an obvious alternative is to apply a method
designed for binary treatment $K$ times. In \S\ref{supple:k-application} of the Supplementary Material we report results from applying the maximum likelihood estimator of \cite{Richardson2017} and the doubly-robust g-estimator of \cite{Dukes2018} twice to estimate $\alpha_1$ and $\alpha_2$ in Setting II, and 
compared their performance to that of the method
in \S \ref{sec:gop}. To apply these methods, we use the subset of units with $Z \in \{0,1\}$
to estimate $\alpha_1$ and the subset with $Z \in \{0,2\}$ to estimate $\alpha_2$.
{As expected, the method of 
	\cite{Richardson2017}
	yields biased estimates as the odds product models are misspecified.}
Similar to the performance reported in Table \ref{tab:sim-all}, two applications of the doubly-robust g-estimator
by \cite{Dukes2018} yield results that are consistent but not efficient; See Table \ref{tab:sim-gop} in the Supplementary Material for details. 
	We use the following generating model:  Treatment $Z$ is assigned according to a multinomial logistic regression model such that 
\begin{align}
	\label{eqn:ps}
	\log \left\{\frac{\pr(Z=1\mid V)}{\pr(Z=0\mid V)} \right\}&= \eta_1^\T V,\quad\hbox{ and }\quad
	\log \left\{\frac{\pr(Z=2\mid V)}{\pr(Z=0\mid V)} \right\}\;=\; \eta_2^\T V,
\end{align}
where $\eta_1  = (1, -1)^\T$, $\eta_2 = (1, -2)^\T$. 
The covariate vector $V$ includes an intercept and a draw from a uniform distribution on $[-2,2]$.

We consider two different settings for the outcome $Y$:
In Setting I, we treat $Z$ as continuous and $Y$ is generated according to models \eqref{eq:model of rr in mono} and \eqref{eq:model of op in mono}, where $\gamma = (0,1)^\T$, $\beta=(1, -0.5)^\T$, so that the log relative risk is linear in $z$. We apply the method of \S \ref{sec:mono} to estimate the relative risk in this setting, and 
compare it to the doubly robust g-estimator of \cite{Dukes2018}.
In Setting II, $Z$ is viewed as unordered categorical 
and the outcome $Y$ is generated from models \eqref{eq:gop-rr} and \eqref{eq:gop-gop}, where $\alpha_1=\gamma=(0,1)^\T$, $\alpha_2=(0, 2)^\T$, $\beta=(1, -0.5)^\T$. Here, we  apply the method of \S \ref{sec:gop} only.


Table \ref{tab:sim-all} summarizes the simulation results for sample size $500$. 
The bias of our proposed estimators is small when the sample size is 500, and further decreases as the sample size increases; see Table \ref{tab:sim-mono}  in the Supplementary Material. The standard deviation accuracy, defined as the ratio of estimated standard deviation and  Monte Carlo standard deviation, is close to $1$ for our proposed estimators. 
The coverage probability of the proposed Wald-type confidence intervals also achieve the nominal 95\% coverage-rate.  Even though, in theory, the doubly robust g-estimator is consistent in this setting as the propensity score model is correctly specified, as shown in Table \ref{tab:sim-all}, with a small sample of 500 the bias can be very large relative to the standard error. In this case, the model-based standard deviation estimate is also much smaller than the Monte Carlo standard deviation. 
One can also see that in this simulation, the proposed estimator is much more efficient than the g-estimator. {In \S\ref{supple:simu-mis-op} of the Supplementary Material, we consider an alternative setting where the odds product model \eqref{eq:model of op in mono} is mis-specified. In this case, when the sample size is large enough, the doubly robust g-estimator of \cite{Dukes2018} has small bias and nominal coverage while the proposed estimator does not.
	However, in the same setting, when the sample size is small, the doubly robust g-estimator has large bias, relative to the  method proposed in \S\ref{sec:mono}}.

{We further note that our method in \S\ref{sec:mono} assumes that treatment effects are monotonic in $z$ for all levels of $v$. In \S\ref{supple:violation-monotonicity} of the Supplementary Material, we conduct additional simulations to evaluate the performance of the method of \S \ref{sec:mono} when this assumption is violated. }

With a categorical treatment taking $K$ + 1 levels an obvious alternative is to apply a method
designed for binary treatment $K$ times. In \S\ref{supple:k-application} of the Supplementary Material we report results from applying the maximum likelihood estimator of \cite{Richardson2017} and the doubly-robust g-estimator of \cite{Dukes2018} twice to estimate $\alpha_1$ and $\alpha_2$ in Setting II, and 
compared their performance to that of the method
in \S \ref{sec:gop}. To apply these methods, we use the subset of units with $Z \in \{0,1\}$
to estimate $\alpha_1$ and the subset with $Z \in \{0,2\}$ to estimate $\alpha_2$.
{As expected, the method of 
	\cite{Richardson2017}
	yields biased estimates as the odds product models are misspecified.}
Similar to the performance reported in Table \ref{tab:sim-all}, two applications of the doubly-robust g-estimator
by \cite{Dukes2018} yield results that are consistent but not efficient; See Table \ref{tab:sim-gop} in the Supplementary Material for details. 

	\section{Application to Titanic Data}\label{sec:application}

We illustrate the use of our proposed methods in \S\ref{sec:method} by studying the association between the passenger class and death in the tragic sinking of the Titanic in 1912. 
We compare the results from our proposed models with those obtained from a generalized linear model.
The data set consists of $1309$ passengers from three passenger classes, of whom  $809$ lost their lives during the event. 
For illustration we removed the $263\ (20.1\%)$ passengers for whom age was missing, resulting in a sample size of $1046$, including $284\ (27.1\%)$ passengers in the first class, $261\ (25.0\%)$ in the second class, and $501\ (47.9\%)$ in the third class. A sensitivity analysis imputing the missing ages provides similar results.
The empirical probability of death is lowest in the first class at $36.3\%$, increasing to  $55.9\%$ in the second class, and $73.9\%$ in the third class. 
Given this, we initially considered modeling the relative risk of death as a monotone function of passenger class, using the first class as the baseline. 

Figure \ref{fig:emp-death} in the Supplementary Material shows the survival statuses of passengers by their passenger class, age and sex. Female passengers tend to have a lower probability of death compared to males, and children tend to have a lower probability of death compared to adults. These observations suggest that the relative risk of death with respect to passenger class may vary with sex and age. 
We let the covariates $X$ and $W$ be identical, which include age, sex, age squared, and the interaction between age and sex. 

We applied five different models to estimate  the variation in the relative risk of death stratifying on age and sex: 1) Poisson regression; 2) Logistic regression; 3) Two applications of the doubly robust g-estimator by \cite{Dukes2018}, where we include the first and second  class passengers in the first application, and the first and third class passengers in the second application; 4) Monotone: the model given by \eqref{eq:model of rr in mono} and \eqref{eq:model of op in mono}; 5) GOP: the model given by \eqref{eq:gop-rr} and \eqref{eq:gop-gop}.  Results for models 1) and 2) were obtained using the {\tt glm} function in {\tt R}  via maximum likelihood with robust standard errors. 

\begin{table}
	\centering
	\begin{threeparttable}
	\def~{\hphantom{$-$}}
	\caption{Coefficient estimates via different models}{%
	\begin{tabular}{l@{ }c@{\hskip 0.05in}c@{\hskip 0.05in }c@{ \hskip 0.05in}c@{ \hskip 0.05in}c@{ \hskip 0.05in}c@{\hskip 0.05in }c@{ \hskip 0.05in}c@{ \hskip 0.05in}c@{\hskip 0.05in }c}
		\hline
		& {\footnotesize 2nd} & {\footnotesize2nd*} & {\footnotesize2nd*} &{\footnotesize2nd*}& {\footnotesize 2nd*} & {\footnotesize 3rd} & {\footnotesize3rd*} & {\footnotesize3rd*} & {\footnotesize3rd*} & {\footnotesize3rd*}  \\ 
		&  & {\footnotesize male} & {\footnotesize age/10} &{\footnotesize age$^2/$}& {\footnotesize male*} &  & {\footnotesize male} & {\footnotesize age/10} & {\footnotesize age$^2/$} & {\footnotesize male*}  \\ 
		& &  &  & 100 & {\footnotesize age/10} &  & & & 100 & {\footnotesize age/10}  \\
		\hline
			\multicolumn{2}{l}{\it Point Estimate}\\[2pt]
			 Monotone & ~1.891 & $-$1.543 & $-$0.165 & ~0.011 & ~0.058 & 3.782 & $-$3.086 & $-$0.329 & 0.022 & ~0.116 \\ 
			GOP & $-$1.134 & ~1.439 & ~0.780 & $-$0.033 & $-$0.617 & 2.204 & $-$1.212 & ~0.053 & 0.020 & $-$0.309 \\ 
			Poisson & $-$1.211 & ~0.938 & ~0.969 & $-$0.072 & $-$0.487 & 2.232 & $-$1.444 & ~0.120 & 0.005 & $-$0.254 \\[6pt] 
			\hline
			\multicolumn{2}{l}{\it Standard Deviation}\\[2pt]
			Monotone & ~0.396 & ~0.407 & ~0.124 & ~0.010 & ~0.107 & 0.792 & ~0.813 & ~0.247 & 0.020 & ~0.214 \\ 
			GOP & ~1.230 & ~1.251 & ~0.369 & ~0.029 & ~0.314 & 0.888 & ~0.957 & ~0.260 & 0.021 & ~0.236 \\ 
			   Poisson & ~2.077 & ~1.967 & ~0.620 & ~0.033 & ~0.542 & 1.874 & ~1.739 & ~0.570 & 0.030 & ~0.482\\[3pt]
			 \hline
	\end{tabular}}
	\begin{tablenotes}
		\item 1st, 2nd, 3rd: the first passenger class, the second passenger class, and the third passenger class.
		The first class is chosen as the baseline. 
	\end{tablenotes}
	\label{tab: estimation-different-model}
	\end{threeparttable}
\end{table}

Table \ref{tab: estimation-different-model} reports regression coefficient estimates from our proposed methods and Poisson regression. {Coefficient estimates for logistic regression are not included here as logistic regression does not directly describe the dependence of the relative risk of death on age and sex so the coefficients are not comparable to those from the other methods.
	The doubly robust g-estimation method 
	did not converge for this model, though it
	did succeed in fitting a simpler model depending solely on the main effects of age and sex.}
The point estimates from the proposed GOP model are close to those from the Poisson model, while the standard errors are smaller.
On the other hand, the point estimates for our Monotone model are different from those given by the other two models. 
Although it appears reasonable from the marginal death rates in each passenger class, the monotonic treatment effects assumption is probably violated after stratifying by age and sex. 
For example, for males from 25 to 57 years old, the empirical probability of death is $62.8\%$ for the first class, $93.0\%$ for the second class, and $82.9\%$ for the third class. 

\begin{figure}[!htbp]
	\centering
	\includegraphics[width=0.9\textwidth]{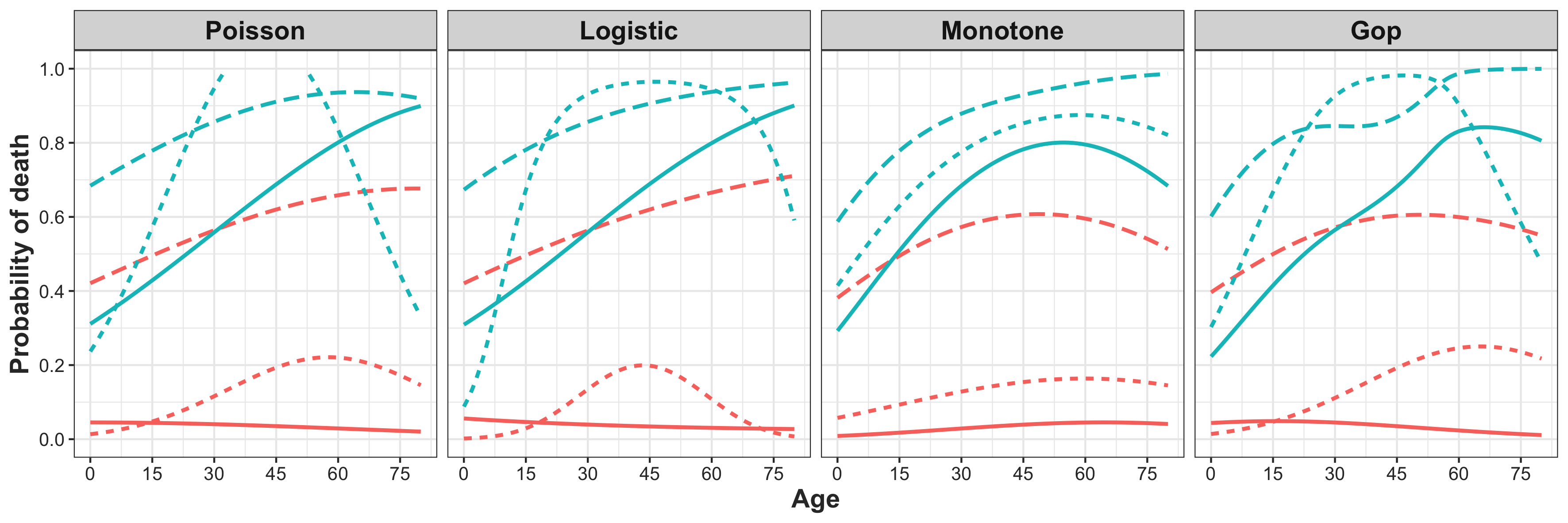}
	\caption{Predicted probability of death of the first passenger class (solid line), the second class (dotted line), and the third class (dashed line) with respect to different models. Red represents female, and blue represents male. }
	\label{fig:pr-by-methods}
\end{figure}

Figure \ref{fig:pr-by-methods} displays the fitted probabilities of death from the Poisson, logistic, Monotone, and Generalized Odds Product models. 
For male passengers in the second class aged between 30 and 50, the fitted probability of death using the Poisson model does not lie in the valid range $[0,1]$.
Under the logistic regression model the fitted probability for second class females decreases to zero as age approaches 80, whereas this does not happen with the Generalized Odds Product model.
However, in the data set, there were only two females in the second class who were older than 55 and both of them died. This suggests that our Generalized Odds Product model may fit the data better.
Unlike the other three plots, the fitted lines from the Monotone model do not cross each other. This is due to the assumption of monotonic treatment effects. As we discussed earlier, this assumption may not be plausible for the Titanic data set.

{The fitted probabilities of death from the two applications of the doubly robust g-estimator are shown separately in Figure \ref{fig:drgest_preds_by_iteration} in the Supplementary Material since each application has a distinct baseline prediction. As explained previously, here we only include the main effects of age and sex.
	In both applications, fitted probabilities of death are greater than 1 for some ages, except for first class females in the second application.} 

\section{Discussion}
We have proposed two novel methods to model multiplicative treatment effects with a binary outcome. 
Our methods can be used either in a likelihood approach or in combination with g-estimation to construct a doubly-robust estimator. 
Such an estimator requires correct modeling of either the  odds products or the propensity score, which is the conditional probability of treatment given covariates. 
More details are given in the Supplementary Material.

\bibliographystyle{apalike}
\bibliography{main}

\newpage
\setcounter{page}{1}

\appendix
\numberwithin{equation}{section}
\numberwithin{table}{section}
\numberwithin{figure}{section}
\begin{center} 
	\LARGE{Supplementary Material for ``Multiplicative Effect Modeling: The General Case''}\\
\end{center}

\begin{abstract}
	In this Supplement, we prove theorems stated in  ``Multiplicative Effect Modeling: The General Case''.  We also give explicit variance formulas for the proposed estimators in \S 2. Furthermore, additional simulation and data application results  are given. Finally, we combine our proposed estimator with g-estimation to construct a doubly-robust estimator.
\end{abstract}

\section{Proof of Theorem 1}\label{appendix:thm-proof}
To prove the existence of a unique set of proper probability distributions $\{p_z(v); z\in \cZ, v\in \cV\}$, it is sufficient to show that  $p_z(v)$ can be written as a function of $h(z, v)$ and $g(v)$.
Since for any $v\in\cV$, $h(z,v)$ is bounded in $z$ and monotonic in $z$, $\lim_{z\rightarrow z_{\inf}} h(z,v)$  and $\lim_{z\rightarrow z_{\sup}} h(z,v)$ exist, denoted as $h_1(v)$ and $h_2(v)$. Without loss of generality, we assume $h(z,v)$ is monotonically non-decreasing in $z$. For simplicity, we denote these by $\lim_{z\rightarrow z_{\inf}}p_{z}(v)$ and $\lim_{z\rightarrow z_{\sup}}p_{z}(v)$ as $p_{z_{\inf}}(v)$ and  $p_{z_{\sup}}(v)$, respectively; let $\Delta(v)=e^{2g(v)}\left(e^{h_1(v)-h_2(v)}+1\right)^2 + 4e^{h_1(v)-h_2(v)+g(v)}\left(1-e^{g(v)}\right)>0$.

For any fixed $v\in\cV$, $ p_{z_{\sup}}(v), p_{z_{\inf}}(v),p_{z_0}(v)$ and $p_z(v)$ satisfy
\begin{align}
	p_{z_{\sup}}(v) &=
	\begin{cases}\frac{e^{g(v)}\{1+e^{h_1(v)-h_2(v)}\}-\sqrt{\Delta(v)}}{2e^{h_1(v)-h_2(v)}\{e^{g(v)}-1\}} & g(v) \neq 0,\\
		\frac{1}{1+e^{h_1(v)-h_2(v)}} & g(v) = 0,
	\end{cases}\\
	p_{z_{\inf}}(v) &= p_{z_{\sup}}(v)e^{ h_1(v)-h_2(v)},\\
	p_{z_0}(v) &= p_{z_{\sup}}(v) e^{-h_2 (v)},\label{eq:p_0}\\
	p_z(v)&=p_{z_{\sup}}(v)e^{h(z,v)-h_2 (v)}\label{eq:p_z},\quad (z\in \cZ).
\end{align}

We now show 
\begin{align}
	\log\{\rr(z_0,z;v)\}&=h(z,v),\label{eq:logrr = h}\\
	\log\{\op(z_{\inf}, z_{\sup};v\} &= g(v).\label{eq:logop = h}
\end{align}
In the case where $g(v)=0$, it is easy to see that  \eqref{eq:logrr = h} and \eqref{eq:logop = h} hold. If $g(v)\neq 0$, for any $v\in \cV$, one may divide \eqref{eq:p_z} by \eqref{eq:p_0} and take the logarithm of both sides. The resulting expression  satisfies \eqref{eq:logrr = h}.
Next we prove that $p_{z_{\sup}}(v)\in (0,1)$, which is equivalent to showing that $p_{z_{\sup}}(v)\{p_{z_{\sup}}(v)-1\}<0$ for any fixed $v$.
\begin{flalign*}
	&{p_{z_{\sup}}(v)\{p_{z_{\sup}}(v)-1\}} \\
	&\quad=\; \frac{\left[e^{g(v)}\{1+e^{h_1(v)-h_2(v)}\}-\sqrt{\Delta(v)}\right]
		\left[e^{g(v)}-e^{h_1(v)-h_2(v)+g(v)}+2e^{
			h_1(v)-h_2(v)}-\sqrt{\Delta(v)}\right]}{\left[2e^{h_1(v)-h_2(v)}\{e^{g(v)}-1\}\right]^2}.
\end{flalign*}
It is enough to prove that the numerator of the above equation is smaller than 0, which can be directly computed. Further $\op(z_{\inf},z_{\sup};v)$ maybe obtained explicitly as:
\begin{align*}
	\MoveEqLeft{\dfrac{p_{z_{\sup}}(v)p_{z_{\inf}}(v)}{\{1-p_{z_{\sup}}(v)\}\{1-p_{z_{\inf}}(v)\}}}\\
	&=\dfrac{\{e^{g(v)}(1+e^{h_1(v)-h_2(v)})-\sqrt{\Delta(v)}\}^2}{\left(e^{h_1(v)-h_2(v)+g(v)}-2e^{h_1(v)-h_2(v)}-e^{g(v)}+\sqrt{\Delta(V)}\right)\left(e^{g(v)}-e^{h_1(v)-h_2(v)+g(v)}-2+\sqrt{\Delta(V)}\right)}\\
	&=\dfrac{e^{g(v)}\left[2e^{g(v)}\left\{e^{h_1(v)-h_2(v)}+1\right\}^2-
		4e^{h_1(v)-h_2(v)}\{e^{g(v)}-1\}-2\{1+e^{h_1(v)-h_2(v)}\}\sqrt{\Delta(v)}\right]}
	{2e^{g(v)}\left\{e^{h_1(v)-h_2(v)}+1\right\}^2-
		4e^{h_1(v)-h_2(v)}\{e^{g(v)}-1\}-2\{1+e^{h_1(v)-h_2(v)}\}\sqrt{\Delta(v)}}\\
	&=e^{g(v)}.
\end{align*}
Thus \eqref{eq:logop = h} is satisfied. 
This completes our proof.

\section{Proof of Theorem 2}\label{appendix:thm2-proof}

In order to prove the map given by
\begin{equation*}
	\left(p_0(v),\ldots,p_K(v)\right) \rightarrow \left(\log \rr(v;0,1), \ldots, \log \rr(v;0,K),\log \gop(v)\right)
\end{equation*}
is a diffeomorphism, we need to prove that for any fixed $v$ and any vector $\left(\rr(0,1;v), \ldots, \rr(0,K;v),\gop(v)\right)\in (\R^+)^{K+1}$, there is one and only one  vector $(p_0(v),\ldots,p_K(v))\in (0,1)^{K+1}$. 
Let $\rr(0,k;v)=c_k(v)\in \R^+$ where $k=1,\ldots, K$, and $\gop(v)=c_{K+1}(v)\in \R^+$. By definition of $\gop(v)$, we further have 
\begin{align}\label{eq:solve p0}
	\MoveEqLeft{\log\{c_{K+1}(v)\}=(K+1)\log\{p_0(v)\}+\sum_{k=1}^K\log\{c_k(v)\} - \log\{1-p_0(v)\}}\\[-2pt]
	&\kern80pt  -\sum_{k=1}^K\log\{1-p_0(v)c_k(v)\}.\nonumber
\end{align}
In the following, we show that there is one and only one solution of Equation \eqref{eq:solve p0} for $p_0(v)\in (0,1)$. For notational simplicity, write $p_k(v)$ as $p_k$, and $c_k(v)$ as $c_k$, $k=0,1,\ldots,K+1$. Let $f(p_0)=(K+1)\log p_0+\sum_{k=1}^K\log(c_k) - \log(1-p_0)-\sum_{k=1}^K\log(1-p_0c_k)-\log(c_{K+1})$.
Now
\begin{align*}
	\frac{df(p_0)}{dp_0} &= \frac{K+1}{p_0}+\frac{1}{1-p_0}+\sum_{k=1}^K \frac{c_k}{1-p_0c_k}\\
	&=\frac{K+1}{p_0}+\frac{1}{1-p_0}+\sum_{k=1}^K \frac{c_k}{1-p_k} > 0.
\end{align*}
Therefore $f(p_0)$ is monotonically increasing on $(0,1)$. Because  $\lim_{p_0\rightarrow 0} f(p_0) = -\infty$ and $\lim_{p_0\rightarrow 1} f(p_0) = +\infty$, there is one and only one root for $f(p_0)=0$ on $(0,1)$. Since the domain of $\mathcal M$, $(\R^+)^{K+1}$, is the Cartesian product of the marginal domains of the Relative Risk and Generalized Odds Product models, the models in $\mathcal M$ are variation independent.



\section{Variance Formulae for the Proposed Estimators}
\subsection*{Estimator Assuming Monotonic Treatment Effect} \label{supp-sec: var mono}
The log-likelihood for a unit can be written as 
\begin{equation}\label{eq:log-likehood in mono}
l(\gamma,\beta|z,v,y) =  y\log\{p_{z}(v;\gamma, \beta)\} +  (1-y)\log\{1-p_{z}(v;\gamma, \beta)\}.
\end{equation}
Without loss of generality, let both the treatment $z_{\min}$ and the baseline treatment be zero. Denote  $\theta(v) = \gamma^\T v$, $g(v) = \beta^\T v$, $\psi(v) = \log p_0(v)$, and  $p_z(v) = e^{z\theta(v) + \psi(v)}\ (z\in \cZ)$. For simplicity, we write $l, \theta, g, \psi, p_z, p_0$ referring to $l(\gamma,\beta|z,v,y),\theta(v), g(v), \psi(v), p_z(v),p_0(v)$, respectively.
The functional dependence structure of the variables  is shown in Figure \ref{fig:var-structure-monon}. Further we have the derivatives of $l(\gamma,\beta|z_i,v_i,y_i)$ with respect to $\gamma$ and $\beta$:
\begin{align}
	\frac{\partial l}{\partial \gamma} &= \frac{\partial l}{\partial p_z}\left( \frac{\partial p_z}{\partial\theta}\frac{\partial \theta}{\partial\gamma} + \frac{\partial p_z}{\partial\psi} \frac{\partial\psi}{\partial \theta}\frac{\partial \theta}{\partial\gamma}\right),\label{eq:l-devarite-alpha}\\
	\frac{\partial l}{\partial \beta} &=  \frac{\partial l}{\partial p_z} \frac{\partial p_z}{\partial\psi}\frac{\partial\psi}{\partial g}\frac{\partial g}{\partial\beta}.\label{eq:l-devarite-beta}
\end{align}
\begin{figure}
	\centering
	\begin{tikzpicture}
	\matrix (m) [matrix of math nodes,row sep=1.5em,column sep=2em,minimum width=1em]
	{
		& & & g & \beta \\
		& & \psi & &  \\
		l & p_z & & \theta & \\
		& & \theta & & \gamma  \\};
	\path[-stealth]
	(m-1-5) edge (m-1-4)
	(m-1-4) edge (m-2-3)
	(m-2-3) edge (m-3-2)
	(m-3-2) edge (m-3-1)
	(m-4-5) edge (m-4-3) edge (m-3-4)
	(m-3-4) edge (m-2-3)
	(m-4-3) edge (m-3-2);
	\end{tikzpicture}
	\caption{Variable structure of the proposed method under the monotonic treatment effects assumption.}
	\label{fig:var-structure-monon}
\end{figure}
In the following, we calculate the terms in \eqref{eq:l-devarite-alpha} and \eqref{eq:l-devarite-beta}. 
\begin{itemize}
	\item[] \begin{equation*}
		\frac{\partial l}{\partial p_z} = \frac{y-p_z}{p_z(1-p_z)},
	\end{equation*}\label{eq:l-pi}
	\item[] \begin{equation*}
		\frac{\partial p_z}{\partial\theta} = zp_z, \quad
		\frac{\partial p_z}{\partial\psi} = p_z.
	\end{equation*}
	\item[] To get $\frac{\partial \psi}{\partial g}, \frac{\partial \psi}{\partial \theta}$, we start from $g(v) = \log \{\op(0,z_{\max})\}$.
	\begin{align*}
		g &=\log \frac{p_0p_{z_{\max}}}{\{1-p_0\}\{1-p_{z_{\max}}\}}\\
		&=\log \frac{p_0^2e^{z_{\max}\theta}}{(1-p_0)(1-p_0e^{z_{\max}\theta})}\\
		&=2\log p_0 + k\theta -\log (1-p_0) - \log (1-p_0e^{z_{\max}\theta})\\
		&= 2\psi + z_{\max}\theta -\log (1-e^{\psi}) - \log (1-e^{\psi+z_{\max}\theta}).
	\end{align*}
	Because $\frac{\partial g}{\partial\theta} = 0$, we further have 
	\begin{align*}
		\frac{\partial g}{\partial\theta} &= 2\frac{\partial \psi}{\partial \theta}+ z_{\max}+ \frac{e^{\psi} \frac{\partial \psi}{\partial \theta}}{1-e^{\psi}} + \frac{e^{\psi+z_{\max}\theta} (\frac{\partial \psi}{\partial \theta}+z_{\max})}{1-e^{\psi+z_{\max}\theta}}\\
		&=2\frac{\partial \psi}{\partial \theta} + z_{\max} + \frac{p_0\frac{\partial \psi}{\partial \theta}}{1-p_0}+\frac{p_{z_{\max}}(\frac{\partial \psi}{\partial \theta}+z_{\max})}{1-p_{z_{\max}}}\\
		&=0.
	\end{align*}
	Therefore, we have 
	\begin{equation*}
		\frac{\partial \psi}{\partial \theta} = -\frac{z_{\max}(1-p_0)}{1-p_0+1-p_{z_{\max}}};
	\end{equation*}
	\begin{align*}
		\frac{\partial g}{\partial\psi}&=2 + \frac{e^\psi}{1-e^\psi}+\frac{e^{\psi + z_{\max}\theta}}{1-e^{\psi + z_{\max}\theta}}\\
		&=2 + \frac{p_0}{1-p_0} + \frac{p_{z_{\max}}}{1-p_{z_{\max}}}.
	\end{align*}
	Then 
	\begin{equation*}
		\frac{\partial \psi}{\partial g} = \frac{(1-p_{z_{\max}})(1-p_0)}{(1-p_{z_{\max}})+(1-p_0)}.
	\end{equation*}
	\item[] We also have
	\begin{equation*}
		\frac{\partial \theta}{\partial \gamma} = v,\quad
		\frac{\partial g}{\partial \beta} = v.
	\end{equation*}
\end{itemize}
With the above building blocks, we finally have the derivatives:
\begin{align}
	\dfrac{\partial l}{\partial \gamma} &=\dfrac{y-p_z}{1-p_z}\cdot\left\{z-
	\dfrac{z_{\max}(1-p_0)}{(1-p_0)+(1-p_{z_{\max}})}\right\}\cdot v,\label{eq：mono, l-to-gamma}\\
	\frac{\partial l}{\partial \beta} &=  \dfrac{y-p_z}{1-p_z} \cdot
	\dfrac{(1-p_0)(1-p_{z_{\max}})}{(1-p_0)+(1-p_{z_{\max}})}\cdot v.\label{eq:mono, l-to-beta}
\end{align}
The Fisher Information matrix $\mathcal I(\alpha, \beta)$ may be calculated to be
\begin{align*}
	\mathcal I(\alpha, \beta)=\mathbb E\left[\left\{\left(\frac{\partial l}{\partial \gamma}\right)^\T, \left(\frac{\partial l}{\partial \beta}\right)^\T\right\}^\T\left\{\left(\frac{\partial l}{\partial \alpha}\right)^\T, \left(\frac{\partial l}{\partial \beta}\right)^\T\right\}\right] &= \mathbb E
	\begin{bmatrix}
		\left(\frac{\partial l}{\partial \gamma}\right)\left(\frac{\partial l}{\partial \gamma}\right)^\T & \left(\frac{\partial l}{\partial \gamma}\right)\left(\frac{\partial l}{\partial \beta}\right)^\T\\
		\left(\frac{\partial l}{\partial \beta}\right)\left(\frac{\partial l}{\partial \gamma}\right)^\T& \left(\frac{\partial l}{\partial \beta}\right)\left(\frac{\partial l}{\partial \beta}\right)^\T
	\end{bmatrix}.
\end{align*}
Then variance covariance matrix for $(\gamma^\T, \beta^\T)^\T$ is $\left\{n\mathcal I(\gamma^\T, \beta^\T)\right\}^{-1}$, where $n$ is the sample size.

·
\subsection*{Estimator Assuming a Categorical Treatment}
Suppose we observe a unit in treatment arm $z_k$. Let $\theta_k = \alpha_k^\T v$.
Then the first derivatives of $l(\alpha_1, \ldots, 
\alpha_K,\beta \mid z, v, y)$ with respect to $\alpha_1,\ldots, \alpha_K, \beta$ are
\begin{align}
	\frac{\partial l}{\partial \alpha_j} &= \frac{y}
	{p_k}\frac{\partial p_k}{\partial \alpha_j} -  \frac{1-y}{1-p_k}\frac{\partial p_k}{\partial \alpha_j} = \frac{y-p_k}{p_k(1-p_k)}\frac{\partial p_k}{\partial \alpha_j} \quad (k = 0,1,\ldots,K; j=1,\ldots,K\label{eq:l-to-alpha_j}),\\
	\frac{\partial l}{\partial \beta} &= \frac{y-p_k}{p_k(1-p_k)}\frac{\partial p_k}{\partial \beta}\label{eq:l-to-beta}.
\end{align}
Since $\partial p_k/\partial \alpha_j=\partial (p_0e^{\theta_k})/\partial \alpha_j$, we further have
\begin{align}
	\frac{\partial p_k}{\partial \alpha_j}&=\left\{
	\begin{array}{ll}
		\frac{\partial p_0}{\partial \alpha_j} e^{\theta_k} & k \neq 0, k\neq j, \\
		\frac{\partial p_0}{\partial \alpha_j}e^{\theta_j} + p_jv & k \neq 0, k= j, \\
		\frac{\partial p_0}{\partial \alpha_j} & k=0;\\
	\end{array}\label{eq:pk-to-alpha_j}
	\right. \\
	\frac{\partial p_k}{\partial \beta}&=\frac{\partial p_0}{\partial \beta} e^{\theta_k}\label{eq:pk-to-beta}.
\end{align}
In order to calculate Eq. \eqref{eq:l-to-alpha_j} and \eqref{eq:l-to-beta}, we need to have $\frac{\partial p_0}{\partial \alpha_j}$ and $\frac{\partial p_0}{\partial \beta}$.
By definition we have
\begin{equation*}
	e^\phi = \frac{\prod _{k=0}^K p_k}{\prod _{k=0}^K (1-p_k)}.
\end{equation*}
Taking the logarithm of both sides gives
\begin{equation}\label{eq:supp-phi-p}
\phi = \sum_{k=0}^K \log p_k - \sum_{k=0}^K \log (1-p_k).
\end{equation}
The derivatives of both sides of \eqref{eq:supp-phi-p} with respect to $\alpha_j$ and $\beta$, respectively, are: 
\begin{align}
	0 &= \dfrac{1}{p_0}\dfrac{\partial p_0}{\partial \alpha_j}\left(\sum_{k=0}^K \dfrac{1}{1-p_j}\right)+\dfrac{1}{1-p_j}v, \label{eq:supp-0-alphaj}\\
	v &= \dfrac{1}{p_0}\dfrac{\partial p_0}{\partial \beta}\left(\sum_{k=0}^K \dfrac{1}{1-p_j}\right). \label{eq:supp-0-beta}
\end{align}
By \eqref{eq:supp-0-alphaj} and \eqref{eq:supp-0-beta}, we further have
\begin{align}
	\dfrac{\partial p_0}{\partial \alpha_j} &= -\dfrac{v\cdot\frac{p_0}{1-p_j}}{\sum_{k=0}^K \frac{1}{1-p_j}},\label{eq:supp-p0-alphaj}\\
	\dfrac{\partial p_0}{\partial \beta} &= \dfrac{p_0v}{\sum_{k=0}^K \frac{1}{1-p_j}}\label{eq:supp-p0-beta}. 
\end{align}
Substituting \eqref{eq:supp-p0-alphaj} and \eqref{eq:supp-p0-beta} into \eqref{eq:l-to-alpha_j} to \eqref{eq:pk-to-beta}, we have
\begin{align*}
	\frac{\partial l}{\partial \alpha_j}&=\left\{
	\begin{array}{ll}
		\frac{v(y-p_k)}{1-p_k}\frac{-\frac{1}{1-p_j}}{\sum_{l=0}^K\frac{1}{1-p_l}} & k\neq 0, k\neq j, \\[2pt]
		\frac{v(y-p_k)}{1-p_k}\left(1-\frac{\frac{1}{1-p_j}}{\sum_{l=0}^K\frac{1}{1-p_l}}\right) &  \quad k\neq 0, k= j, \\[2pt]
		\frac{v(y-p_k)}{1-p_k}\frac{-\frac{1}{1-p_j}}{\sum_{l=0}^K\frac{1}{1-p_l}} & k=0;
	\end{array}\label{eq:pi-to-alpha_j 2}
	\right.\\[4pt]
	\frac{\partial l}{\partial \beta}&=\frac{(y-p_k)v}{1-p_k}\frac{1}{\sum_{l=0}^K\frac{1}{1-p_l}}.
\end{align*}
The variance-covariance matrix for $(\alpha_1,\ldots,\alpha_K,\beta)$ can be calculated as the inverse of the Fisher Information matrix.

In general, the likelihood is not concave. In practice, we use a simple iterative procedure for finding a solution to the score equations. To be more specific: for the method which assumes monotonicity, we assign a starting value for $\gamma$ and $\beta$. At each step $t$, we first find $\gamma^{(t)}$ via maximizing the (profile) log-likelihood while holding $\beta$ fixed at $\beta^{(t-1)}$; we then find the  optimal $\beta^{(t)}$ via maximizing the  log-likelihood holding 
$\gamma$ fixed at $\gamma^{(t)}$. The iterations stop when the differences between the parameters at successive iterations are smaller than a pre-defined threshold. 

Similarly, for the Generalized Odds Product method we optimize the parameters of one of the models $\alpha_1,\ldots ,\alpha_K,\beta$ while holding fixed the parameters of the other models.

\section{Additional Simulations}

\subsection{Simulation results for the setting of Table \ref{tab:sim-all} and $n=1000, 5000$}

Table \ref{tab:sim-mono} summarizes the simulation results corresponding to Table \ref{tab:sim-all} in the main paper for  sample sizes 1000 and 5000.

\begin{table}[!htbp]
	\begin{threeparttable}
	\def~{\hphantom{$-$}}
	\caption{Simulation results for three different methods based on 1000, 5000 samples and 1000 Monte Carlo runs. The true values for $\gamma$, $\alpha_1$ and $\alpha_2$ are $(0,1)^\T$, $(0,1)^\T$ and $(0,2)^\T$ respectively}{%
	\begin{tabular}{lcccccccccc}
		\\[-4pt]
		
		& \multicolumn{2}{c}{Bias${\scriptscriptstyle \times 10^2}$(Standard Error${\scriptscriptstyle \times 10^2}$)} && \multicolumn{2}{c}{SD Accuracy} && \multicolumn{4}{c}{Coverage  (Nominal = 95\%)}\\[5pt] 
		\midrule
		$n=1000$ & \multicolumn{2}{c}{$\gamma$}   & & \multicolumn{2}{c}{$\gamma$}   &&&\multicolumn{2}{c}{$\gamma$}&\\[3pt]
		Setting I &\multicolumn{6}{l}{Monotone}\\[3pt]
		& \multicolumn{2}{c}{$-$0.122(0.145)}  && \multicolumn{2}{c}{1.020} &&&\multicolumn{2}{c}{0.957}&  \\ 
		& \multicolumn{2}{c}{~0.011(0.174)} &&   \multicolumn{2}{c}{1.025} &&& \multicolumn{2}{c}{0.954}&\\
		Setting I & \multicolumn{6}{l}{DR-G}\\[3pt]
		& \multicolumn{2}{c}{$-$0.363(0.391)}  && \multicolumn{2}{c}{0.827} &&& \multicolumn{2}{c}{0.924} & \\ 
		& \multicolumn{2}{c}{~9.178(0.653)} &&  \multicolumn{2}{c}{0.713} &&& \multicolumn{2}{c}{0.883} &\\[5pt]
		& $\alpha_1$ & $\alpha_2$ && $\alpha_1$& $\alpha_2$ &&& $\alpha_1$ & $\alpha_2$ &\\ 
		Setting II &  \multicolumn{6}{l}{GOP}\\[3pt]
		& 0.252(0.346) & 0.231(0.346) && 1.008 & 0.999 &&& 0.951 & 0.950 \\ 
		& 0.291(0.417) & 0.598(0.395) && 0.997 & 0.999 &&& 0.948 & 0.950 \\ 
		[8pt]
		\midrule
		$n=5000$ & \multicolumn{2}{c}{$\gamma$}    && \multicolumn{2}{c}{$\gamma$}  &&& \multicolumn{2}{c}{$\gamma$} \\
		Setting I &\multicolumn{6}{l}{Monotone}\\[3pt]
		&  \multicolumn{2}{c}{0.002(0.064)} && \multicolumn{2}{c}{1.032} &&&    \multicolumn{2}{c}{0.956}  \\ 
		& \multicolumn{2}{c}{0.068(0.079)}   && \multicolumn{2}{c}{1.001}   &&& \multicolumn{2}{c}{0.956}  \\ 
		Setting I & \multicolumn{6}{l}{DR-G}\\[3pt]
		& \multicolumn{2}{c}{$-$0.026(0.138)} && \multicolumn{2}{c}{0.959} &&& \multicolumn{2}{c}{0.937} \\ 
		& \multicolumn{2}{c}{~1.290(0.212)} &&  \multicolumn{2}{c}{0.914}  &&& \multicolumn{2}{c}{0.927}  \\ [5pt]
		& $\alpha_1$ & $\alpha_2$ && $\alpha_1$& $\alpha_2$ &&& $\alpha_1$ & $\alpha_2$\\ 
		Setting II &  \multicolumn{6}{l}{GOP}\\[3pt]
		& $-$0.417(0.151) & $-$0.264(0.145) && 1.017 & 1.048 &&& 0.950 & 0.961 \\ 
		& $-$0.323(0.186) & ~0.033(0.177) && 0.987 & 0.982 &&& 0.954 & 0.947 \\ 
		
\end{tabular}}
	\label{tab:sim-mono}
	\begin{tablenotes}
		\item Monotone, using models \eqref{eq:model of rr in mono} and \eqref{eq:model of op in mono}; DR-G, doubly robust estimator by \cite{Dukes2018}; GOP, using models  \eqref{eq:gop-rr} and \eqref{eq:gop-gop}.
		\item SD Accuracy = estimated standard deviation / Monte Carlo standard deviation.
	\end{tablenotes}
\end{threeparttable}
\end{table}

\subsection{Simulations with a mis-specified  odds product model}
\label{supple:simu-mis-op}

Here we consider the same data generating model as in Section 3, Setting I. An analyst takes variable $V^*$ instead of $V$ in the nuisance model to estimate parameters of interest; here $V^*$ includes an intercept and another covariate which is a transformation of $V$, specifically $2\cos(V)$. 
Results are shown in Table \ref{tab:sim-miss-gop1}.
As expected,  when the sample size is large enough, the doubly robust g-estimator of \cite{Dukes2018} has small bias and nominal coverage while the proposed estimator does not.
When the sample size is small, however, the doubly robust g-estimator can have large bias, relative to our proposed Monotonic method.

\begin{table}[!htbp]
	\begin{threeparttable}
	\def~{\hphantom{$-$}}
	\caption{Simulation results based on 500, 1000, and 5000 samples and 1000 Monte Carlo runs for the relative risk model with an odds product nuisance model (Setting I). The nuisance model is misspecified. The true value for  $\gamma$ is $(0,1)^\T$}{%
\begin{tabular}{lcccccccc}
	\\[-4pt]
	& Bias${\scriptscriptstyle \times 10^2}$(Standard Error${\scriptscriptstyle \times 10^2}$) && SD Accuracy && Coverage (Nominal = 95\%)\\[5pt]
	\midrule
	$n=500$ &  $\gamma$ && $\gamma$ && $\gamma$ \\ [3pt]
	& \multicolumn{5}{l}{Monotone}\\[3pt]
	& ~0.857(0.219) && 0.958 && 0.948 \\ 
	& $-$0.669(0.25) && 1.033 && 0.953 \\ 
	& \multicolumn{5}{l}{DR-G}\\[3pt]
	& $-$0.521(0.924) && 0.542 && 0.891 \\ 
	& ~22.525(1.477) && 0.472 && 0.797 \\ 
	[8pt]
	\midrule
	$n=1000$ &  $\gamma$ && $\gamma$ && $\gamma$ \\ [3pt]
	& \multicolumn{5}{l}{Monotone}\\[3pt]
	& ~0.770(0.146) && 1.002 && 0.955 \\ 
	& $-$0.443(0.181) && 0.990 && 0.946 \\ 
	& \multicolumn{5}{l}{DR-G}\\[3pt]
	& $-$0.498(0.393) && 0.818 && 0.933 \\ 
	& ~8.372(0.636) && 0.730 && 0.874 \\  
	[8pt]
	\midrule
	$n=5000$ &  $\gamma$ && $\gamma$ && $\gamma$ \\ [3pt]
	& \multicolumn{5}{l}{Monotone}\\[3pt]
	& ~0.705(0.064) && 1.019 && 0.948 \\ 
	& $-$0.921(0.078) && 1.016 && 0.945 \\ 
	& \multicolumn{5}{l}{DR-G}\\[3pt]
	& 0.116(0.130) && 1.011 && 0.958 \\ 
	& 1.253(0.203) && 0.953 && 0.945 \\   
\end{tabular}}
	\label{tab:sim-miss-gop1}
	\begin{tablenotes}
		\item Monotone: Using models \eqref{eq:model of rr in mono} and \eqref{eq:model of op in mono}; DR-G, doubly robust estimator by \cite{Dukes2018}.
		\item SD Accuracy = estimated standard deviation / Monte Carlo standard deviation.
	\end{tablenotes}
\end{threeparttable}
\end{table}

\subsection{Sensitivity of the Monotone model to violations of the monotonicity assumption}
\label{supple:violation-monotonicity}

We consider a setting where the treatment $Z$ is a draw from the uniform distribution on $\{0,1,2\}$ and the covariate vector $V$ includes an intercepts and a draw from the uniform distribution on $[-2,2]$. The outcome $Y$ is generated according to models \eqref{eq:model of rr in mono} and \eqref{eq:model of op in mono}, except with outcome probabilities swapped between $Z=1$ and $Z=2$ for certain units randomly picked from the sample, resulting in a
violation of monotonicity  for those samples. The results are given in Table \ref{tab:sim-non-mono}. As expected,
the bias increases and the coverage decreases with the proportion of  the sample for which the monotonicity assumption is violated. 

\begin{table}[!htbp]
	\def~{\hphantom{$-$}}
	\caption{Simulation results for the Monotone model with partially non-monotonic data based on 1000 samples and 1000 Monte Carlo runs. The true values for $\gamma$ and $\beta$ are $(0, 1)^\T$ and $(-0.5, 1)^\T$ respectively}{%
	\begin{tabular}{lccc}
		\\[-4pt]
		$\%$ Monotonic & Bias${\scriptscriptstyle \times 10^2}$(Standard Error${\scriptscriptstyle \times 10^2}$) & SD Accuracy & Coverage  (Nominal = 95\%)\\[3pt] 
		\midrule
		& $\gamma$ & $\gamma$ & $\gamma$ \\
		[3pt]
		\multirow{2}{4em}{75$\%$} & $-$0.079(0.002) & 1.079 & 0.713\\
		& $-$0.263(0.002) & 1.021 & 0.004 \\
		[5pt]
		\multirow{2}{4em}{90$\%$} & $-$0.027(0.002) & 1.051 & 0.937\\
		& $-$0.127(0.002) & 1.042 & 0.420 \\
		[5pt]
		\multirow{2}{4em}{100$\%$} & $-$0.002(0.002) & 0.999 & 0.947\\
		& 0.006(0.002) & 0.970 & 0.949 \\
\end{tabular}}
	\label{tab:sim-non-mono}
\end{table}

\subsection{Comparison with $K$-applications of methods designed for binary treatments}
\label{supple:k-application}

With a categorical treatment taking $K+1$ levels, a naive alternative is to use $K$ applications of a method designed for modeling the relative risk for a binary treatment. 
In this case we compare our proposed generalized odds product method to two previously proposed relative risk models for binary treatment:  the likelihood method proposed by \cite{Richardson2017} and the doubly robust g-estimator of \cite{Dukes2018}.   To apply these methods, we use the subset of units with $Z\in \{0,1\}$ to estimate $\alpha_1$ and the subset with $Z\in \{0,2\}$ to estimate $\alpha_2$.  For the method of \cite{Richardson2017}, we assume that 
\[
\log \op(0,1; v) = \frac{p_0(v) p_1(v)}{(1-p_0(v))(1-p_1(v))} = \beta_1^\T v;\quad \log \op(0,2; v) = \beta_2^\T v.
\]
In general, these odds product models will be incompatible with the models for $\rr(0,1; v), \rr(0,2; v)$ as they are variation dependent. For 
the method of \cite{Dukes2018}, we  assume the propensity score model in \eqref{eqn:ps}, and a baseline model $E(Y\mid V, A=0) = \exp(\xi^\T V)$.

The outcome $Y$ is generated from models \eqref{eq:gop-rr} and \eqref{eq:gop-gop}, where the true values for  $\alpha_1$, $\alpha_2$, $\beta$ are $(-0.5,1)^\T$, $(0.5,1.5)^\T$, and $(1,-0.5)^\T$, respectively.
Table \ref{tab:sim-gop} shows the simulation results for sample sizes 500, 1000 and 5000. The biases of our point and variance estimators are small and  go to zero as the sample size increases. Although the bias of the doubly robust g-estimator is large at $n=500$, the bias decreases with the sample size. {The biases for two applications of the  likelihood method of
	\cite{Richardson2017} are relatively small, suggesting that the odds product models are not severely mis-specified in this case.} 



\begin{table}[!htbp]
	\centering
	\begin{threeparttable}
	\def~{\hphantom{$-$}}
	\caption{Simulation results for three different methods based on 500, 1000 and 5000 samples and 1000 Monte Carlo runs. The true values for $\alpha_1$ and $\alpha_2$ are $(-0.5, 1)^\T$ and $(0.5, 1.5)^\T$ respectively}{%
\begin{tabular}{lcccccccccc}
	& \multicolumn{2}{c}{Bias${\scriptscriptstyle \times 10^2}$(Standard Error${\scriptscriptstyle \times 10^2}$)} && \multicolumn{2}{c}{SD Accuracy} && \multicolumn{4}{c}{Coverage (Nominal = 95\%)}\\[5pt] 
	\midrule
	$n=500$ &  $\alpha_1$ & $\alpha_2$ && $\alpha_1$ & $\alpha_2$ &&& $\alpha_1$ & $\alpha_2$\\
	& \multicolumn{6}{l}{GOP}\\
	& ~0.612(0.626) & ~1.735(0.451) && 1.007 & 1.008 &&& 0.957 & 0.960 \\ 
	& ~0.080(0.669) & $-$0.490(0.463) && 1.011 & 0.988 &&& 0.961 & 0.951 \\ 
	[3pt]
	&     \multicolumn{6}{l}{DR-G (applied twice)}\\
	& $-$8.251(0.988) & ~3.030(0.565) && 0.828 & 0.939 &&& 0.922 & 0.947 \\ 
	& ~20.80(1.943) & ~14.98(1.224) && 0.653 & 0.695 &&& 0.885 & 0.933 \\ 
	[3pt]
	&    \multicolumn{6}{l}{OP (applied twice)}\\
	& ~0.146(0.638) & ~2.129(0.457) && 0.998 & 1.011 &&& 0.961 & 0.959 \\ 
	& ~1.913(0.764) & $-$1.618(0.47) && 0.971 & 1.008 &&& 0.956 & 0.955 \\ 
	[8pt]
	\midrule
	$n=1000$ &  $\alpha_1$ & $\alpha_2$ && $\alpha_1$ & $\alpha_2$ &&& $\alpha_1$ & $\alpha_2$\\
	& \multicolumn{6}{l}{GOP}\\
	& $-$0.433(0.434) & ~0.626(0.314) && 1.005 & 0.996 &&& 0.964 & 0.946 \\ 
	& ~0.519(0.456) & $-$0.124(0.314) && 1.026 & 0.989 &&& 0.963 & 0.952 \\ 
	[3pt]
	&   \multicolumn{6}{l}{DR-G (applied twice)}\\[3pt]
	& $-$4.705(0.605) & ~1.020(0.363) && 0.878 & 0.962 &&& 0.936 & 0.947 \\ 
	& ~9.834(1.143) & ~6.266(0.715) && 0.733 & 0.722 &&& 0.911 & 0.947 \\ 
	[3pt]
	&   \multicolumn{6}{l}{OP (applied twice)}\\[3pt]
	& $-$0.531(0.443) & ~1.106(0.320) && 1.002 & 0.994 &&& 0.963 & 0.952 \\ 
	& ~1.112(0.523) & $-$1.251(0.321) && 0.991 & 1.002 &&& 0.956 & 0.947 \\ 
	[8pt]
	\midrule
	$n=5000$ &  $\alpha_1$ & $\alpha_2$ && $\alpha_1$ & $\alpha_2$ &&& $\alpha_1$ & $\alpha_2$\\
	&\multicolumn{6}{l}{GOP}\\
	& ~0.006(0.189) & ~0.122(0.134) && 1.014 & 1.027 &&& 0.959 & 0.956 \\ 
	& $-$0.261(0.203) & ~0.099(0.136) && 1.016 & 1.00 &&& 0.953 & 0.964 \\ 
	[3pt]
	&    \multicolumn{6}{l}{DR-G (applied twice)}\\[3pt]
	& $-$0.566(0.216) & ~0.091(0.144) && 1.027 & 1.026 &&& 0.958 & 0.948 \\ 
	& ~1.120(0.357) & ~1.219(0.210) && 0.972 & 0.963 &&& 0.949 & 0.942 \\ 
	[3pt]
	&   \multicolumn{6}{l}{OP (applied twice)}\\[3pt]
	& $-$0.002(0.194) & ~0.462(0.136) && 1.009 & 1.027 &&& 0.953 & 0.961 \\ 
	& $-$0.203(0.230) & $-$0.798(0.142) && 0.992 & 0.984 &&& 0.949 & 0.956 \\
\end{tabular}}
	\label{tab:sim-gop}
	\begin{tablenotes}
		\item GOP: Using models \eqref{eq:gop-rr} and \eqref{eq:gop-gop}; DR-G, doubly robust estimator by \cite{Dukes2018};  OP: Using nuisance model proposed by \cite{Richardson2017}.
		\item SD Accuracy = estimated standard deviation / Monte Carlo standard deviation.
	\end{tablenotes}
\end{threeparttable}
\end{table}

\section{Additional results for the data application}

\begin{figure}[h]
	\centering
	\includegraphics[width=.8\textwidth]{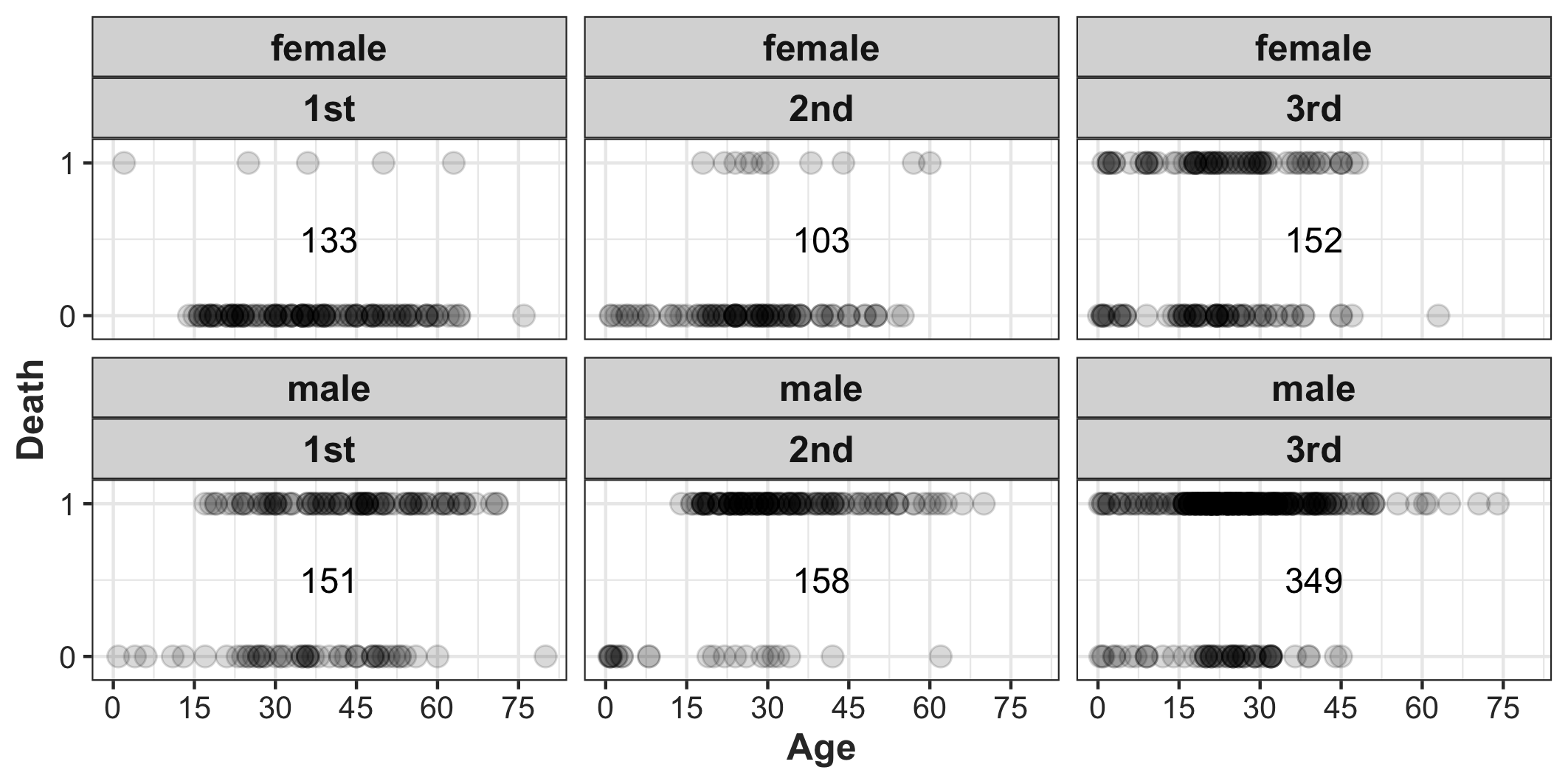}
	\caption{Passengers' survival statuses by passenger class, age, and sex. The number of passengers in each group is shown in the center of the corresponding plot. 
	}
	\label{fig:emp-death}
\end{figure}

\begin{figure}[!htbp]
	\centering
	\includegraphics[width=0.9\textwidth]{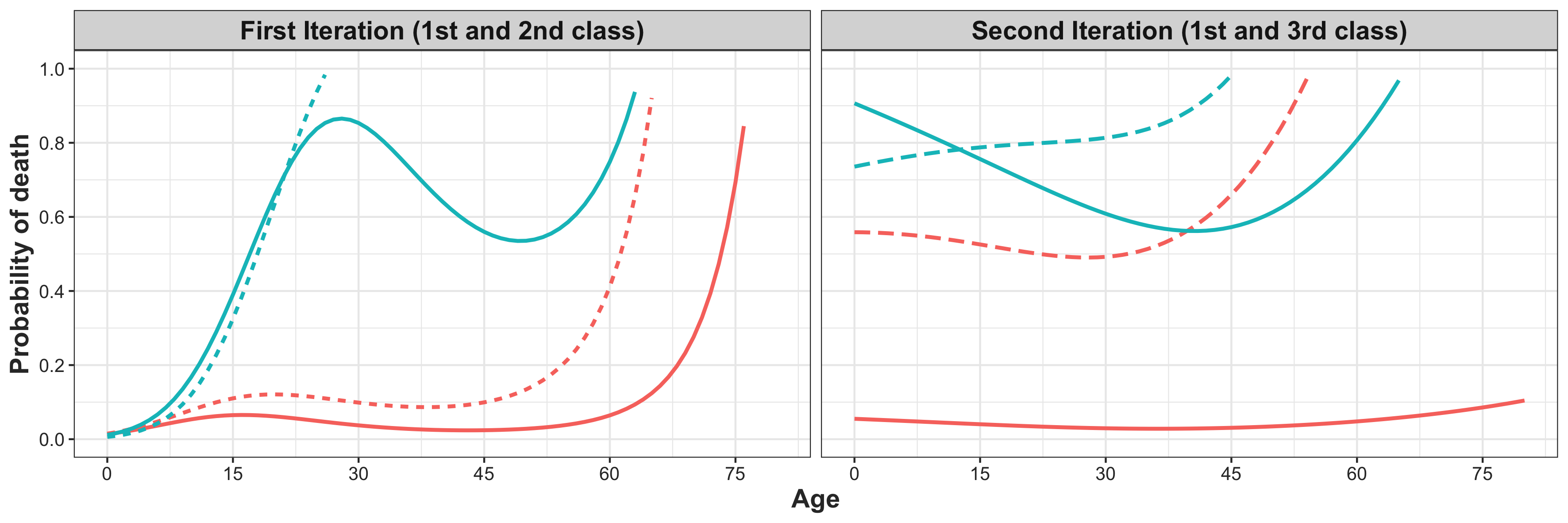}
	\caption{Predicted probability of death of the first passenger class (solid line), the second class (dotted line), and the third class (dashed line) with respect to applications of the doubly robust g-estimation method of \cite{Dukes2018}. Red represents female, and blue represents male. }
	\label{fig:drgest_preds_by_iteration}
\end{figure}

\section{Doubly robust estimator}
\citet[][\S A.15]{van2011targeted} have provided the efficient scores for the parameters of interest in  relative risk models which allow the treatment to be categorical or continuous, and also allow interactions between treatment and baseline covariates. In the following, we separately show the score functions for our two proposed methods.
\begin{itemize}
	\item Parameterization assuming a monotonic relative risk. Our model of interest is $\log \{\rr(0,Z;V,\gamma)\}=\gamma ^\T VZ$. The efficient score function is 
	\begin{equation}
	S(\gamma) = \left\{Y\exp(-\gamma ^\T VZ) - p_0(V)\right\}h(Z\mid V),
	\end{equation}
	where 
	\begin{equation}
	h(Z\mid V) = \frac{Vp_Z(V)}{p_0(V)\{1-p_Z(V)\}}\left[Z - \frac{E\left\{\frac{Zp_Z(V)}{1-p_Z(V)} \mid V\right\}}{E\left\{\frac{p_Z(V)}{1-p_Z(V)} \mid V\right\}}\right].
	\end{equation}
	This representation shows $ES(\gamma)=0$ if either the model for the baseline $p_0(V)$ is correctly specified or the conditional distribution of treatment $Z$ given covariates $V$ is correctly specified. This yields a doubly robust estimator for $\gamma$.\\
	\item Parameterization with a categorical treatment (without a monotonicity assumption). The model of interest is $\log \{\rr(0,Z;V)\} = \sum_{k=1}^K \mathbbm{1}\{Z=k\}\alpha_k^\T V$. Let $S(\alpha) = \left(S(\alpha_1)^\T,\ldots,S(\alpha_K)^\T\right)$ be the score function for $(\alpha_1,\ldots, \alpha_K)$. Similarly to the monotonic treatment effect model, 
	\begin{equation}
	S(\alpha_i) = \left[Y\exp\left\{-\sum_{k=1}^K \mathbbm{1}\{Z=k\}\alpha_k^\T V\right\} - p_0(V)\right]h_i(Z\mid V)\quad i\in \{1,\ldots, K\},
	\end{equation}
	where
	\begin{equation}
	h_i(Z\mid V) =  \frac{Vp_Z(V)}{p_0(V)\{1-p_Z(V)\}}\left[\mathbbm{1}\{Z=i\}-\frac{E\left\{\frac{[\mathbbm{1}\{Z=i\}p_Z(V)}{1-p_Z(V)} \mid V\right\}}{E\left\{\frac{p_Z(V)}{1-p_Z(V)} \mid V\right\}}\right].
	\end{equation}
	As in our first method, the doubly robust estimator of $(\alpha_1,\ldots, \alpha_K)$ can be shown to be consistent if either the baseline risk model or the conditional probability distribution $\pr(Z\mid V)$ are correctly specified.
\end{itemize}

\end{document}